\documentclass[aps,pre,twocolumn,superscriptaddress,amsmath,amssymb,longbibliography]{revtex4-1}

\usepackage{graphicx}
\usepackage{dcolumn}
\usepackage{bm}
\usepackage{hyperref}
\usepackage{comment}
\usepackage{color}
\usepackage{soul}
\begin{document}

\preprint{APS/123-QED}

\title{Failure processes of cemented granular materials}

\author{Yuta Yamaguchi}%
\email{yamaguchi@eps.s.u-tokyo.ac.jp}
\affiliation{%
 Department of Earth and Planetary Science, University of Tokyo, Tokyo 113-0033, Japan.}
 \affiliation{%
Department of Earth and Space Science, Osaka University, Osaka 560-0043, Japan.}%
\author{Soumyajyoti Biswas}%
\affiliation{%
WW8-Materials Simulation, Department of Materials Science, Friedrich-Alexander-Universit\"at Erlangen-N\"urnberg, F\"urth 90762, Germany.}%
\affiliation{%
Department of Physics, SRM University - AP, Andhra Pradesh, India}%
\author{Takahiro Hatano}%
\affiliation{%
Department of Earth and Space Science, Osaka University, Osaka 560-0043, Japan.}%
\author{Lucas Goehring}%
\email{lucas.goehring@ntu.ac.uk}
\affiliation{%
School of Science and Technology, Nottingham Trent University, Nottingham NG11 8NS, UK.}%

\date{\today}

\begin{abstract}
The mechanics of cohesive or cemented granular materials is complex, combining the heterogeneous responses of granular media, like force chains, with clearly defined material properties.  Here, we use a discrete element model (DEM) simulation, consisting of an assemblage of elastic particles connected by softer but breakable elastic bonds, to explore how this class of material deforms and fails under uniaxial compression.  We are particularly interested in the connection between the microscopic interactions among the grains or particles and the macroscopic material response.  To this end, the properties of the particles and the stiffness of the bonds are matched to experimental measurements of a cohesive granular media with tunable elasticity.  The criterion for breaking a bond is also based on an explicit Griffith energy balance, with realistic surface energies.  By varying the initial volume fraction of the particle assembles we show that this simple model reproduces a wide range of experimental behaviors, both in the elastic limit and beyond it.  These include quantitative details of the distinct failure modes of shear-banding, ductile failure and compaction banding or anti-cracks, as well as the transitions between these modes.  The present work, therefore, provides a unified framework for understanding the failure of porous materials such as sandstone, marble, powder aggregates, snow and foam.
\end{abstract}
\maketitle

\section{\label{sec:introduction}INTRODUCTION}
A wide variety of materials are made up of individual grains that are held together by a matrix material. Examples include both  naturally occurring materials such as mortars, asphalts, volcanic ashes and snow (see \textit{e.g.}, \cite{jiang2013} and references therein) as well as artificially generated materials such as cemented solids and sintered glass or alumina beads \cite{hemmerle2016,schmeink2017,wang2019}.  
Consequently, the failure of porous materials is also known from many contexts. In geological situations, landslides, snow avalanches, and earthquakes are representative examples. The failure of industrial materials such as foams or ceramics and building materials including sandstone, concrete and marble are widely studied in the materials sciences \cite{lawn}. Given a fixed sample size, the observed modes of failure processes of cohesive granular materials are determined by two basic competing ingredients: stress localization due to damage and scatter of damage due to disorder. When we regard the porosity of the media as a measure of disorder in the system, a transition from brittle to ductile deformation is expected \cite{wong2012} with increasing porosity \textit{i.e.}, decreasing packing fraction of the grains in the material. Further increases in porosity can lead to another localized mode of failure such as an anti-crack, which can be defined as a fracture mode with the displacement field of a mode-I crack, but opposite in sign (\textit{i.e.} compression or closure rather than an opening displacement) \cite{fletcher1981, sternlof2005, heierli2008}. 

Here, we perform discrete element model (DEM) simulations on a model cohesive granular media, designed around the measured inputs from a particularly well-characterized experimental system \cite{hemmerle2016,schmeink2017}. The quantitative verification of our model is done by comparison with experiments in terms of the stress-strain relationship. We find that the system density plays an essential role in the transitions between three distinct failure modes: brittle failure, ductile failure, and compaction banding or anti-cracks. While each of these modes is well-known in different contexts, we show how they can all be captured in a single model framework, and one in which every parameter is constrained by direct experimental measurement.  Specifically, we show how transitions between these modes of fracture can be observed through a variation of the system density or packing fraction. 

In laboratory experiments, the brittle failure of rocks has been studied for predicting the timing of material failure. The nature of the microcrack events, typically detected as acoustic signals in experiments, is one indicator used to predict the failure timing and this method was developed through many laboratory experiments \cite{lockner1991,sammonds1992,olsson2000,zang2000,baud2004,fortin2006,townend2008}. The interplay between the stress localization due to damage and the diffused damage due to disorder in the system can be observed by tracking the spatial locations of the acoustic signals. One can observe that the apparently randomly distributed micro-cracks progressively localize to form a system-spanning crack leading to catastrophic breakdown. Experimental studies  \cite{renard2009, renard2018} have also used X-rays to measure more details of brittle faulting of rocks. Similarly, a conductivity-based method was recently used to detect microcracks in sandstone under strong temperature variations rather than direct mechanical loading \cite{ilin2020}. In general, the statistical properties of the microcrack events, for a range of porosities and materials, have been studied experimentally (see \textit{e.g.} \cite{vives13a,vives13b}) and also numerically using discrete element model (DEM) simulations \cite{kun14,pal2016}. 

The plastic deformation mode of failure is commonly observed in metallic materials but is also seen in rocks under particular conditions such as high confining pressures \cite{karman1911,paterson1958} or high temperature \cite{heard1960}. Lowering strain rate and relatively higher porosity play a role in reducing the transition pressure \cite{karman1911, handin1957, handin1963, robinson1959}, although this porosity dependence has only been observed in highly porous media such as sandstones and similar rocks.  However, the brittle-ductile transition and strain-rate hardening due to interaction and sizes of shear bands are also observed in micrometer scale metals \cite{cynthia1,cynthia2}.

In a different context, pure compressional deformation bands, which develop in a transverse direction to the confining axis, have been reported in geological field studies \cite{hill1989, mollema1996}. Along these lines, earlier work has been done including theoretically modeling such modes \cite{olsson1999, issen2000}, laboratory experiments \cite{olsson1999, olsson2000, wong2001, baud2004, fortin2006, baud2012}, field studies \cite{eichhubl2010, schultz2010, fossen2011}, and simulations \cite{katsman2005, marketos2009}. These compressional bands are called compaction bands or anti-cracks and can also be observed in highly porous materials like foams \cite{reis2009} and snows \cite{kinosita1967, barraclough2017}. We note that this compaction band preferentially develops in low-density systems. After their formations, such bands can work as a barrier to movement or flow, and contribute to the directivity of permeability in sandstone aquifers \cite{baud2012,sternlof2006}, for example.  The growth and extension of compaction bands are regarded as anti-crack formation and can be represented by the Eshelby inclusion model~\cite{sternlof2005}.

Recently, cohesive or cemented granular materials with tunable material properties have been created artificially and the elastic properties of these materials are studied in laboratory experiments which directly motivate our model~\cite{hemmerle2016, schmeink2017}. The cohesive granular media were made from glass beads held together by small amounts of a curable polymer (PDMS). When added as a liquid, the polymer settles into pendular or capillary bridges between nearby particles, which solidify into rigid bonds between the particles after heating. The stiffness of the bonds is much smaller than that of the beads (see Table \ref{tab:parameter_info}), similar to the situation also investigated in Ref.~\cite{wang2019}. Microscopic experiments were performed to investigate the elastic properties of a single bond which connects two glass beads. From such microscopic experiments, the effective spring constants of the polymer bonds are extracted. Based on these information, one can model a cohesive granular media through a DEM \cite{Cundall1979} simulation and reproduce the macroscopic experimental results primarily in the elastic regime, including the effects of bond stiffness \cite{hemmerle2020}. 

In this work, we propose a model that captures the initial elastic and eventual yielding processes in cohesive granular media, for a range of packing fractions, in a systematic way, bridging the similar experimental observations performed for specific ranges of packing fractions. The model is simplified in some aspects (see sec. \ref{sec:method}) to reduce the number of parameters required to describe it. But, the parameters that are present in the model are fully constrained by the inputs from the laboratory experiments (see Table \ref{tab:parameter_info}). Despite its simplicity, we find that the model can reproduce a wide range of experimental and observational behavior of real materials, including sandstone \cite{olsson2000,baud2004,townend2008}, snow \cite{barraclough2017} and foam \cite{reis2009}, in addition to the exemplar material \cite{hemmerle2016,schmeink2017} that it is modeled after. In particular, power law scalings of the elastic modulus and the yield stresses with the effective bond density are shown in this model for a range of packing fractions. Beyond the yield point, the different modes by which a cohesive granular system fails are also seen in the model. For example, our model is able to reproduce the failure modes of cohesive granular media with the variation of the packing fraction--starting from a shear band formation at higher densities, through a ductile failure in the intermediate densities, to a compaction failure at low densities (high porosity). These modes agree with the corresponding experimental observations. For example, in the case of sandstone, compaction band features are known to form preferentially in more porous rock, specifically where the porosity is above about 20\% \cite{mollema1996,olsson2000,townend2008}.  Similarly, our model can reproduce an observed transition towards shear-band formation \cite{baud2005} when the system becomes dense enough to be dilatant.  This indicates that the model captures the essential physics of the modes of deformation of cohesive granular media.  Conversely, it shows how these very different modes of failure can occur in what are otherwise similar materials, in response to relatively small changes in their composition.

These results show that our model, constituted with the microscopic parameter inputs from experiments, reproduces experimentally observed failure and pre-failure responses for a range of packing fractions \textit{i.e.}, from shear banding to ductile failure to anti-cracks. It also reveals the statistical variations in localizations of microcracks and their quantification. The model, therefore, opens up the possibilities for investigations in various parameter ranges and making quantitative predictions that are hard to probe otherwise. 

\section{\label{sec:method}METHODS}

\begin{figure}
    \includegraphics[width=1.\linewidth]{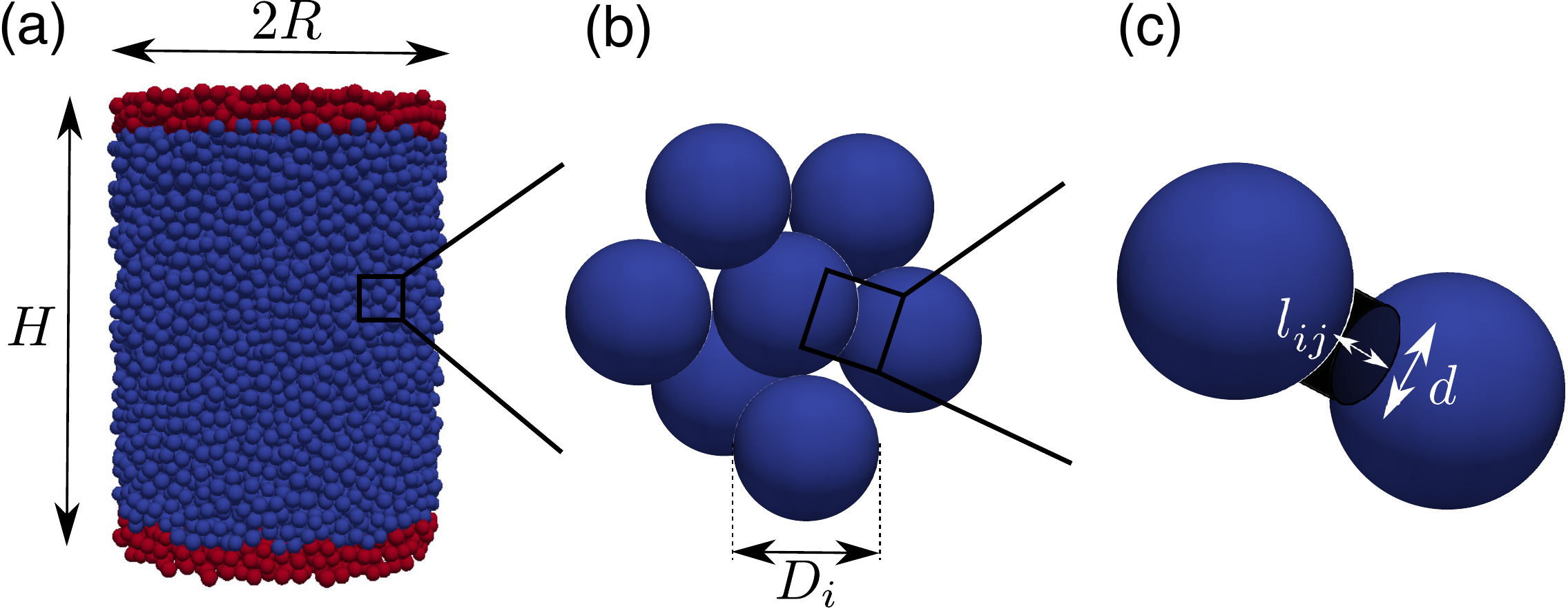}
    \caption{Modeling a cohesive granular medium involves considerations of different scales.  (a) We simulate the compression of a cylinder with radius $R$ and height $H$.  It is composed of spherical particles, grains or beads, held together by elastic bridges.  Particles near the upper and lower boundaries (red) are clamped, so that they can only move vertically.  (b) The magnified view shows randomly arranged and slightly polydisperse particles, where particle $i$ has a diameter $D_i$. Each particle can interact with its neighbors by contact forces, and by any bridges or bonds.  (c) Zooming in more, we model a bond between particles $i$ and $j$ as a truncated cylinder of height $l_{ij}$ and diameter $d$.  This bond can stretch, both normally and tangentially, and will break if strained enough.}
    \label{fig:geometry}
\end{figure}

Our model is motivated by experimental studies of cohesive granular media.  In particular, we focus on a single experimental system that has been particularly well-characterized \cite{hemmerle2016, schmeink2017,hemmerle2020}.  We assume that the macroscopic properties of these materials can be described by the constitutive laws governing the microscopic deformations of individual particles and the fragile cohesive bonds, or bridges, that link them.  Therefore, we have designed DEM simulations where the various elastic responses of the components (\textit{e.g.} spring constants) are constrained by the measurements of micro-mechanical experiments involving two glass beads connected by an isolated polymer bridge~\cite{hemmerle2020}.  Although we focus on information from one particular set of experiments (see Table~\ref{tab:parameter_info}), this modeling method can be extended to any similar cohesive granular material by suitably choosing these material constants.

\begin{table}[]
    \centering
    \setlength{\tabcolsep}{1pt}
    \scalebox{1}[1]{
    \begin{tabular}{|l|c|c|c|c|}\hline
         property                     & symbol           & value  & scaled value& ref. \\ \hline \hline
         sample diameter              & $2R$             & $4.22\ {\rm mm}$    & $2R/D$             & \cite{hemmerle2016} \\
         sample height                & $H$              & $5.91\ {\rm mm}$    & $H/D$              & -- \\ \hline
         bond diameter                & $d$              & $75.4\ {\rm \mu m}$ & $d/D$              & \cite{schmeink2017} \\ 
         bond stiffness, norm.       & $k_n^{\rm bond}$ & $5.76\ {\rm kN/m}$  & $k_n^{\rm bond}/k_n^{\rm glass}$ & \cite{schmeink2017} \\
         bond stiffness, tang.   & $k_t^{\rm bond}$ & $0.32\ {\rm kN/m}$  & $k_t^{\rm bond}/k_n^{\rm glass}$ & \cite{schmeink2017} \\
         friction coefficient & $\mu$           & $0.50$               & $0.50$            & \cite{penskiy2011} \\ \hline
          bead diameter               & $D$              & $200.9\ {\rm \mu m}$ & $1$               & \cite{hemmerle2016} \\
          bead mass                   & $m$              & $10.8\ {\rm \mu g}$  & $1$               & \cite{hemmerle2016} \\ 
         bead stiffness, norm.       & $k_n^{\rm glass}$& $57.6\ {\rm kN/m}$   & $1$               & \cite{hemmerle2020} \\
         bead stiffness, tang.   & $k_t^{\rm glass}$& $3.25\ {\rm kN/m}$   & $k_t^{\rm glass}/k_n^{\rm glass}$  & \cite{hemmerle2020} \\ \hline
         interfacial energy           & $G$              & $7\    {\rm J/m^2}$ & $G/k_n^{\rm glass}$  & \cite{schmeink2017,chopin2011} \\
         \hline     
    \end{tabular}
    }
    \caption{Model properties set to experimentally observed values, corresponding to glass beads held together by softer (PDMS) bridges.  In particular, Refs. \cite{hemmerle2016,schmeink2017,hemmerle2020} correspond to observations made in identically prepared materials.  The sample height $H$ is 1.5 times that studied in Ref. \cite{hemmerle2016}.}
    \label{tab:parameter_info}
\end{table}

To simplify the simulation parameters, we also non-dimensionalize our model using this particular set of experimental parameters \cite{hemmerle2016,schmeink2017,hemmerle2020}. Specifically, we scale the system such that the average mass ($m$) and diameter ($D$) of the particles, along with the spring constant for normal glass-glass contact ($k_n^{\rm glass}$) take on unit values. For this we use an average particle diameter of $200.9\ {\rm \mu m}$ and an average particle mass of $10.8\ {\rm \mu g}$.  Other units are scaled accordingly, for example velocities are scaled by $D(k_n^{\rm glass}/m)^{1/2}$.  A list of the scaled simulation parameters is included in Table \ref{tab:parameter_info}.   The $k_n^{\rm glass}$ is set ten times higher than the normal spring constant of a polymer bridge, where the latter is estimated as $5.76\ {\rm kN/m}$ \cite{hemmerle2020}.  This ratio is low enough to prevent numerical instabilities, yet high enough to capture the difference of stiffness between the glass beads and their softer bonds.  We confirmed this by performing several simulations in which this stiffness ratio was 20, 50 or 100, and in which we did not see any significant change in terms of stress-strain curves.

\subsection{DEM: equations of motion}

We consider the dynamics of a collection of spherical particles that can interact with each other through contact forces and cohesive bonds.  The equations of motion are adapted from studies of granular media, in particular Ref.\ \cite{luding2008}. The particles are indexed such that particle $i$ has diameter $D_i$, mass $m_i$, center position $\bm{r}_i$, angular velocity $\bm{\omega}_i$ and moment of inertia $I_i = m_i D_i^2 / 10$. The equations of motion of any individual particle involve the translational and rotational degrees of freedom, and are described as
\begin{align}
    &m_i\frac{d^2\bm{r}_i}{dt^2}=\sum_{j\neq i}(F_{ij}^n \bm{n}_{ij} + \bm{F}_{ij}^t ),\label{eq:eqm}\\
    &I_i \frac{d\bm{\omega}_i}{dt} = \frac{D_i}{2} \sum_{j \neq i} \bm{n}_{ij}\times \bm{F}_{ij}^t.
    \label{eq:rotation}
\end{align}
The right hand sides of these equations give the inter-particle forces and torques felt between particle $i$ and all other particles.  For an interaction between particles $i$ and $j$ the force components $F_{ij}^n$ and $\bm{F}_{ij}^t$ are defined, respectively, as the projections of the total inter-particle force onto the line connecting the centers of the two particles (\textit{i.e.} with normal unit vector $\bm{n}_{ij}=(\bm{r}_i-\bm{r}_j)/|\bm{r}_i-\bm{r}_j|$), and the plane normal to that line.  

We account for the forces due to the contact of adjacent beads pressing on each other, as well as those arising from the distortion of cohesive bonds.  For both contributions we use a linear elastic response, where the stiffness of the particles is assumed to be much higher than that of the bonds, and include dissipation during particle contact. The interactions depend on the surface separation of the particles, $\delta_{ij}^n = |\bm{r}_i-\bm{r}_j|-(D_i+D_j)/2$, which is negative when particles overlap. 

\begin{figure}
    \includegraphics[width=1.\linewidth]{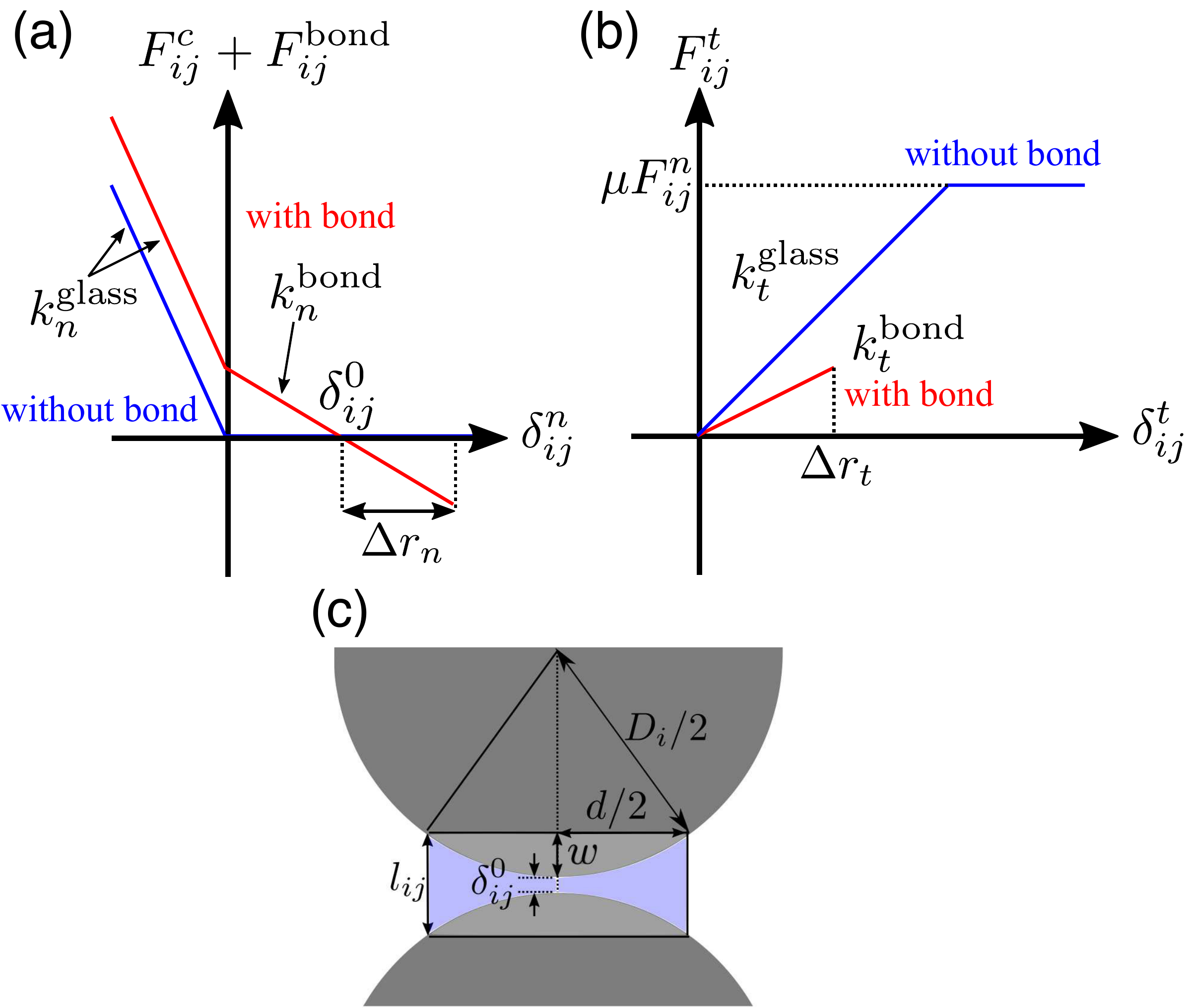}
    \caption{Modeling inter-particle interactions.  (a) The normal forces between two interacting particles, $i$ and $j$, depends on their surface separation, $\delta_{ij}^n$, and whether or not they share an elastic bond. (b) Similarly, their shear force depends on the tangential displacement, $|\bm{\delta}_{ij}^t|$, of any contact or bond. In both panels the red curves show the situation of a shared cohesive bond, whereas the blue curves show the case of no cohesive interaction. (c) A sketch of the geometry of two glass beads held together by a cohesive bond. }
    \label{fig:force}
\end{figure}

The normal component of the interaction has three possible contributions: a repulsive contact force, an elastic restoring force from any cohesive bond and a dissipation force.  As sketched in Fig.~\ref{fig:force}(a), this is summarized as
\begin{equation}
    F_{ij}^n =F_{ij}^c + F_{ij}^{\rm bond} + F_{ij}^{\rm diss}.
\end{equation}
The conservative aspects are given by
\begin{align}
    F_{ij}^c + F_{ij}^{\rm bond}
    = \begin{cases}
    	-k_n^{\rm glass}\delta_{ij}^n + k_n^{\rm bond} \delta_{ij}^0 & \delta_{ij}^n \le 0,\\
	 - k_n^{\rm bond}\left( \delta_{ij}^n - \delta_{ij}^0\right) & \delta_{ij}^n > 0.
    \end{cases}\label{eq:contact_force}
\end{align}
Here, the cohesive bond is modeled as a spring of normal stiffness $k_n^{\rm bond} = 0.1$ and an equilibrium length equal to the initial particle separation, $\delta_{ij}^0$. Its effect is only considered if there is an intact bond linking the particles; otherwise we set $k_n^{\rm bond}=0$.  Particle overlap is also treated as a spring, but of stiffness $k_n^{\rm glass} = 1$.  Additionally, overlap causes dissipation, which is modeled as
\begin{align}
    F_{ij}^{\rm diss}
    = \begin{cases}
    	-\zeta v^n_{ij}& \delta_{ij}^n \le 0,\\
	0 & \delta_{ij}^n > 0,
    \end{cases}\label{eq:dissipation}
\end{align}
where $\zeta$ is a dissipation rate. This ensures that collisions are inelastic.  The dissipation depends on the relative velocity at the contact point, which is computed as
\begin{equation}
    \bm{v}_{ij} =\frac{d\bm{r}_{i}}{dt} - \frac{d\bm{r}_{j}}{dt} - \bm{n}_{ij}\times 
    \frac{1}{2}(D_i\bm{\omega}_i + D_j\bm{\omega}_j).
    \label{relative velocity}
\end{equation}
Note that here, for simplicity, we ignore the effects of any small overlap when computing velocities (\textit{i.e.} in Eq. \ref{relative velocity}, we consider $D_i$ rather than $D_i+ \delta_{ij}^{n}$). As with forces, we project $\bm{v}_{ij}$ onto the normal direction connecting the centers of the interacting particles, ${v}^n_{ij} = \bm{v}_{ij} \cdot \bm{n}_{ij}$, and onto the plane orthogonal to this line, $\bm{v}^t_{ij} = \bm{v}_{ij} - v^n_{ij}\bm{n}_{ij}$.  

Finally, we treat the tangential component of the inter-particle forces, as shown in Fig.~\ref{fig:force}(b).  If two particles are linked by a cohesive bond, and as long as they are not also physically overlapping, then there is a damped, elastic restoring force acting to return them to their original configuration, 
\begin{equation}
     \bm{F}_{ij}^t =  k_t^{\rm bond}\bm{\delta}_{ij}^t - \zeta \bm{v}^t_{ij}.
\end{equation}
In this case $\bm{\delta}^t_{ij}$ is the relative tangential displacement away from the initial contact point, and is measured by integrating $\bm{v}^t_{ij}$ over time.  If two particles are overlapping then there is instead a restoring force
\begin{equation}
     \bm{F}_{ij}^t = -(k_t^{\rm glass}\bm\delta_{ij}^t + \zeta \bm v^t_{ij}).
\end{equation}
This force is limited by Coulomb friction to have a maximum magnitude of $\mu |F^n_{ij}|$, for friction coefficient $\mu$.  Here, $\bm\delta^t_{ij}$ is the tangential displacement measured relative to the point of first contact between the particles. 
   
\subsection{\label{sec:initial_config}Model setup}

\begin{table}[]
    \centering
    \setlength{\tabcolsep}{8pt}
    \begin{tabular}{|c|c|c|c|c|}\hline
         $\phi$ & $N$ & $\sigma(N)$ & $Z$ & $\sigma(Z)$ \\ \hline
         $0.350$  & $6{,}754$   & $38$ & $3.887$ & $0.061$ \\ 
         $0.375$  & $7{,}207$   & $44$ & $4.040$ & $0.061$ \\ 
         $0.400$  & $7{,}721$   & $59$ & $4.242$ & $0.060$ \\ 
         $0.425$  & $8{,}194$   & $36$ & $4.535$ & $0.143$ \\ 
         $0.450$  & $8{,}708$   & $39$ & $4.841$ & $0.088$ \\ 
         $0.475$  & $9{,}189$   & $46$ & $5.218$ & $0.078$ \\ 
         $0.500$  & $9{,}665$   & $19$ & $5.721$ & $0.127$ \\ 
         $0.525$  & $10{,}136$  & $15$ & $6.261$ & $0.028$ \\
         $0.550$  & $10{,}616$  & $18$ & $7.018$ & $0.038$ \\ 
         $0.575$  & $11{,}084$  & $25$ & $7.837$ & $0.020$ \\
         $0.580$  & $11{,}188$  & $10$ & $7.979$ & $0.015$ \\ \hline     
    \end{tabular}
    \caption{In our initial configurations the average number of particles, $N$, and their coordination number, $Z$, depend on the packing fraction, $\phi$. The tabulated values are averaged over five independent realizations with different initial configurations and standard deviations are indicated by $\sigma$.}
    \label{tab:initial_info}
\end{table}

Particles and the cohesive bonds between them determine our system properties.  Here, we describe how we set up their initial configurations (see Table \ref{tab:initial_info}) and boundary conditions.  We first consider a cubic region with periodic boundaries in the horizontal directions and rigid walls at the upper and lower boundaries.  This box is seeded with particles at random positions, with the number of particles ultimately determining the packing fraction $\phi$.  The particles are initially point-like but the particle diameters are slowly increased, and their positions updated to avoid overlaps, following the methods detailed in Refs. \cite{Clarke1987, clarke1993}. During this process we maintain a small polydispersity in particle diameters.
Specifically, the particle diameter distribution is a Gaussian with a standard deviation of $1\% $. We use the Box-Muller method \cite{box1958} to reproduce the polydispersity.

Once a final packing has been made we identify the $N$ particles that are entirely within a cylinder of radius $R$ and height $H$, and remove all outliers.  We then add a cohesive bond between any two particles if their initial surface separation, $\delta_{ij}^n$, is below $0.1\, D$. This range gives us the same density of bonds as was observed in our motivating experiments--specifically, particles form bonds to $Z=7.9$ neighbors, on average, for a packing fraction $\phi=0.59$ \cite{schmeink2017}.  The typical numbers of particles modeled, the coordination numbers of their packings and how these depend on $\phi$ are given in Table\ \ref{tab:initial_info}.  

During deformation we adopt clamped boundary conditions (See Fig.~\ref{fig:geometry}(a)).  For this, all particles within $2.5\, D$ of either the top or bottom edges of the cylinder are constrained: they are only allowed to move vertically and their lateral displacement is prohibited.  The position of the top boundary is then fixed and the bottom one moves upwards at a constant velocity of $10^{-4}$,  corresponding to $46.4\ {\rm mm/s}$. This is expected to be slow enough to match with quasi-static experiments, as we confirmed that results are reproducible with a slower velocity of $10^{-5}$.  At the upper boundary the loading force, $F^{\rm wall}$, is calculated by summing the vertical components of all the forces applied on the boundary particles from the bulk of the sample. The normal stress of the system is then
\begin{equation}
    \sigma = \frac{F^{\rm wall}}{\pi R^2}.
    \label{eq:stress}
\end{equation}
Similarly, the normal strain of the system, $\epsilon$ follows from the displacement of the lower boundary.

In every time step, the model records the positions of the beads and the configuration of the bonds and forces between them.  The number of bonds associated with each bead gives its coordination number $Z_i$, with an average value $Z$ that will tend to decrease over time as damage accumulates by processes which we now describe. 

\subsection{\label{sec:criteria_bond}Breaking cohesive bonds}

A system of particle assemblies with inter-particle bonds or beams requires some criterion for bond breakages. In some previous work this has been based on a von-Mises criterion for yielding \cite{kun14,pal2016} or the Tresca failure criterion for maximum shear stress \cite{brendel2011}. Here, our model for breaking cohesive bonds is based on energetic considerations and is designed around the Griffith criteria for fracture.  
In our motivating experiments, failure happens by bonds detaching or peeling away from particles \cite{schmeink2017}.  Hence, we allow a bond to break when the internal elastic energy stored in its deformation exceeds the peeling energy.

The energy needed to peel a cohesive bond off a particle is proportional to the bond's surface area,
\begin{equation}
	U_s=\frac{\pi d^2}{4} G,
	\label{eq:surface_energy}
\end{equation}
where $G$ is the interfacial energy (\textit{i.e.} the critical strain energy release rate) of the bond.   For this we use $G=7\ {\rm J/m^2}$, as measured for peeling PDMS from glass \cite{chopin2011}.  This value is consistent with the fracture toughness of our experimental cohesive granular materials \cite{schmeink2017}.

The strain energy stored in a bond of volume $V$ under a normal uniaxial strain $\epsilon_{n}$ is modeled as
\begin{eqnarray}
	U^n  =\frac{1}{2}V{\epsilon_n}^2 E_p  =\frac{\pi d^2l_{ij}}{8}\left(\frac{\Delta r_n}{l_{ij}}\right)^2 E_p.
	       \label{eq:normal_strain_energy}
\end{eqnarray}
For this, we regard a cohesive bond as a pillar of height $l_{ij} = 2w + \delta_{ij}^0$ (see Fig.~\ref{fig:force}(c)) and volume $V=\pi d^2 l_{ij}/4$, and take the normal displacement of the bond to be  $\Delta r_n=\delta_{ij}^n - \delta_{ij}^0$.   Likewise, the contribution to the strain energy from any tangential displacement is given by 
 \begin{equation}
 	U^t=\frac{\pi d^2l_{ij}}{8}\left(\frac{|\bm{\delta}_{ij}^t|}{2l_{ij}}\right)^2 E_g,
	\label{eq:tangential_strain_energy}
 \end{equation}
 where $E_g = E_p/(2(1+\nu))$ is the shear modulus of the bond material.  As befits a polymer bond, we use a Poisson ratio $\nu=1/2$, such that $E_g = E_p/3$.
 
A comparison between the surface energy (Eq.\ (\ref{eq:surface_energy})) and the strain energy (sum of Eqs.\ (\ref{eq:normal_strain_energy}) and (\ref{eq:tangential_strain_energy})) now gives us the failure condition
\begin{eqnarray}
	\Delta r_n^2 + \frac{|\bm{\delta}_{ij}^t|^2}{12} \ge 2\frac{G l_{ij}}{E_p}.
	\label{energy_balance}
 \end{eqnarray}
Whenever any bond exceeds this criterion, it is removed from the simulation.  Throughout a numerical experiment we track the locations of the bonds as they break.  These correspond to microcracks or similarly localized damage, events which can be recorded as acoustic signals in laboratory experiments. 

\begin{figure}
    \includegraphics[width=1.\linewidth]{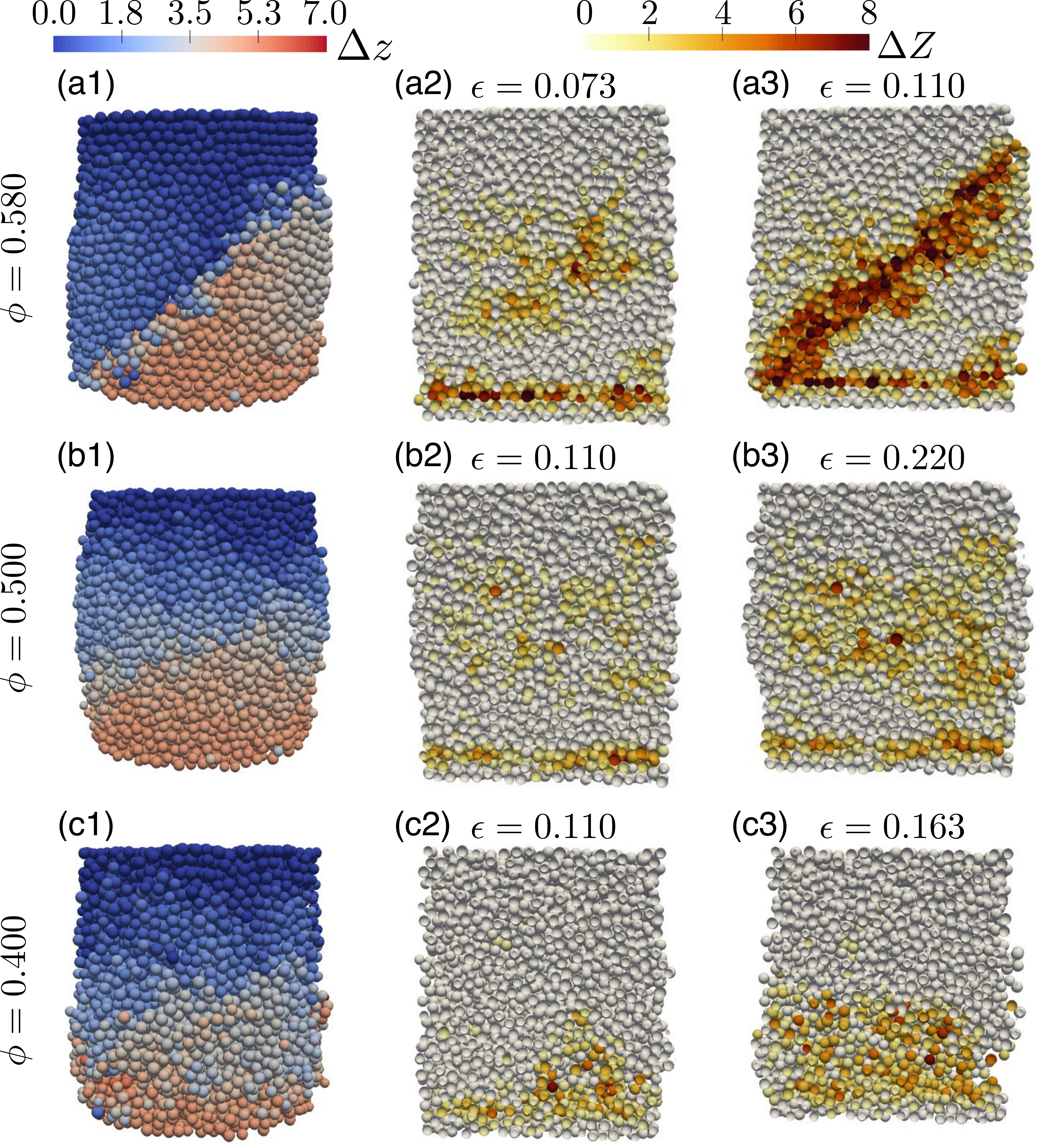}
    \caption{Deformation and failure modes of cohesive granular media at various packing fractions, $\phi$.  Examples are shown, by row, for representative values of (a) a dense packing of $\phi=0.580$, (b) an intermediate case of $\phi = 0.500$ and (c) a low value of $\phi = 0.400$.  In column (1) the particles at the surface of the deformed samples are colored to show their vertical displacements, $\Delta z$, compared to their initial positions.  The sample in (a1) fails along a slip plane or shear band, whereas (b1) shows the yielding of the system by plastic deformation and uniform compaction is seen in (c1). Columns (2) and (3) show cross-sectional snapshots of these processes at various strains, which highlight the changes in coordination number, $\Delta Z$, as the deformation proceeds.  In all cases there is activity at the lower boundary, resulting from the clamped conditions.  Additionally, (a) shows damage localizing along an inclined shear band, whereas (c) highlights a horizontal failure plane.}
    \label{fig:snapshot}
\end{figure}

\section{\label{sec:results}Results and discussion}

We will now explore how the failure of the cohesive granular material under compression is affected by its initial packing fraction, focusing on simulations between $\phi = 0.350$ and $0.580$.  In a related study incorporating experimental findings \cite{hemmerle2020}, we also describe how changing bond stiffness affects the elastic response of this type of material.
The presentation and discussion of results will progress from the elastic response of the system at low strains, through to the statistical features of the micro-crack assembly as yielding begins, and finally to the modes of failure seen at different $\phi$.

An overview of our results is given in Fig.~\ref{fig:snapshot}, which demonstrates how samples at different packing fractions deform and fail.   At very high densities of particles, samples develop shear bands, such as the $\phi=0.580$ example in Fig.~\ref{fig:snapshot}(a).
At an intermediate value of $\phi = 0.500$, as in Fig.~\ref{fig:snapshot}(b), this mode of compressional failure is suppressed, and the sample deforms plastically, showing barreling or swelling around the mid-plane of the sample.  Finally, and representative of low-$\phi$ behavior, Fig.~\ref{fig:snapshot}(c) shows how a $\phi=0.400$ cylinder compresses almost perfectly uniaxially, until failure localizes along a horizontal plane and an anti-crack \cite{reis2009}, or compression failure event \cite{baud2004,townend2008}, passes from right to left across lower part of the sample.   In the following we will discuss these results in more detail, starting with an investigation of the initial elastic regime of this type of material.  

\subsection{\label{sec:elastic_regime}The elastic regime}

\begin{figure*}
    \includegraphics[width=1.0\linewidth]{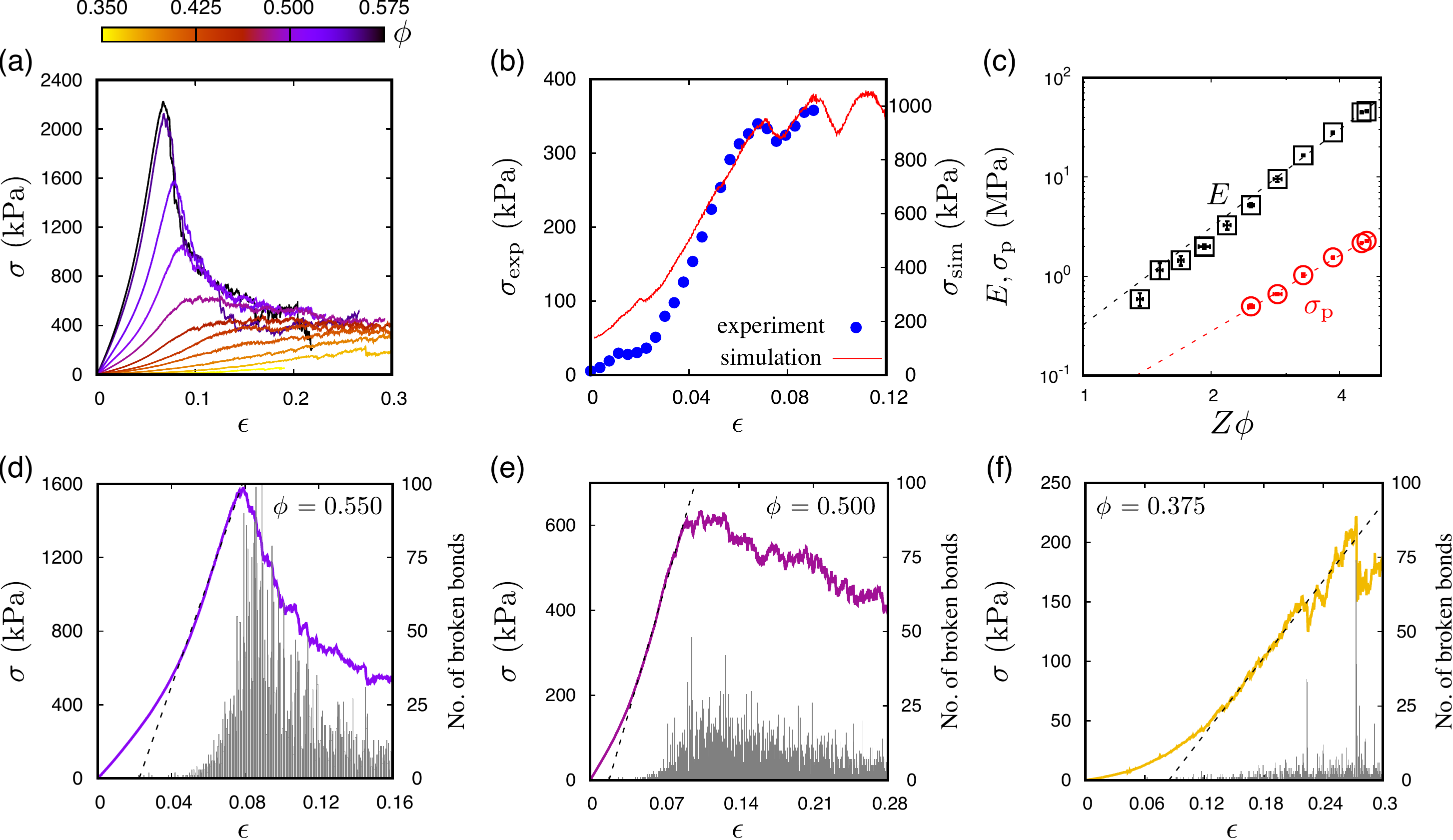}
    \caption{Elastic response of cohesive granular media.  (a)  The stress-strain curves depend on the initial packing fraction, $\phi$. As $\phi$ decreases the material becomes softer, and the  peak stress drops.  The post-peak behavior also changes, from a more brittle to a plastic, yielding response. (b) A comparison can be between an experimental stress-strain curve measured in our motivating material \cite{hemmerle2016} and simulation results where the initial particle positions are taken directly from that experiment.  (c) The Young's modulus (black squares) depends on the initial density of contacts or cohesive bonds, \textit{i.e.} with $Z\phi$, in a manner consistent with a power law of exponent $\xi = 3.27\pm0.14$. This broadly agrees with the scaling reported for other cases of adhesive (sticky) granular media or aggregates \cite{kendall1987,kendall2001,shih1990,mewis2012,gaume2017}. Simultaneously, the peak stress, $\sigma_\mathrm{p}$ (red circles), depends on $Z\phi$, consistent with a power-law fit of exponent $\alpha = 2.53 \pm 0.12$. The global microcrack activity (here, bonds broken in each $3\times 10^{-4}$ strain step) also varies with $\phi$. Representative plots show (d) brittle failure at $\phi=0.550$, with an accompanying burst of activity and a large post-peak stress drop; (e) ductile failures at $\phi=0.500$, with a broader and less-defined peak of activity; and (f) the formation of an anti-crack at $\phi=0.375$, with intermittent stress drops and associated spikes of microcrack activity. The dashed lines in (d)-(f) show fits to the elastic regime, used for estimating $E$.} 
    \label{fig:stress-activity}
\end{figure*}

The bulk response of our model of cohesive granular material to compression is shown in Fig.~\ref{fig:stress-activity}(a), which gives the stress-strain relationships of the uniaxial compression of samples of different initial packing fractions.  For small strains, \textit{i.e.} before the system experiences any significant yielding, the material behaves as an elastic solid.  By fitting the slope of this response, leading up to the peak stress, we measured the Young's modulus, $E$ of each sample. For our highest packing fraction of $\phi=0.580$ the shape of the stress-strain curve closely matches that of the corresponding experiments \cite{hemmerle2016}, from which we took our model parameters, although $E$ is larger than what was measured there. Part of this difference is due to a simplification of the model, which assigns all cohesive bridges the same spring constant, regardless of their initial length.  The model's clamped boundary conditions also contribute to its stiffness, as compared to the experiment.  We tested this by repeating a simulation run without the clamped conditions, and using a shorter cylinder whose dimensions matched the test sample geometry used in Hemmerle \textit{et al.} \cite{hemmerle2016}.  Under these conditions we found $E=13.7$ MPa, compared to the experimental value of $7.9$ MPa \cite{hemmerle2016}. The agreement between experiment and simulation is made increasingly clear if we used the exact particle positions as measured by in the X-ray tomography experiments reported in \cite{hemmerle2016}, while also maintaining the average coordination number measured in those tests \cite{schmeink2017}; the corresponding stress-strain curves are shown in Fig.~\ref{fig:stress-activity}(b). Further quantitative comparisons with experiments are also possible within the elastic regime, starting from the pairwise responses of the beads at the microscopic level, which we intend to explore elsewhere \cite{hemmerle2020}.

Systems with a higher density of particles, and hence a higher coordination number and more bonds, are stiffer than those with a lower $\phi$.   As shown in Fig.~\ref{fig:stress-activity}(c), the Young's modulus increases smoothly with  $Z\phi$, which describes the density of inter-particle contacts before the system deforms.  This dependency is well-fit by a power law, $E \sim (Z\phi)^\xi$, with exponent $\xi= 3.27\pm0.14$.  As we discuss below, this scaling is generally in line with the expected behavior of aggregated systems.  

The stiffness of granular materials, especially those with some form of adhesion or cohesion, is known to depend strongly on the packing fraction and structure of its particles (\textit{e.g.} \cite{kendall1987,kendall2001,shih1990,mewis2012,gaume2017}). The effective-medium theory of a frictional but cohesionless granular media would suggest a linear scaling of $E$ with $Z\phi$ \cite{walton1987}.  However, by independently varying $Z$ and $\phi$ for sticky hard spheres, Gaume \textit{et al.} \cite{gaume2017} showed a data collapse for $E(Z\phi)$, with a power law like ours but with an exponent of $\xi = 4.9$.   Similar results have been reported in 2D simulations of cohesive particles \cite{gilabert2008,roy2016}, with stiffnesses again increasing much faster with $Z\phi$ than would be naively be predicted by effective-medium theory.  This high sensitivity of the bulk stiffness to contact density has been attributed to the presence of force chains, which will generate a backbone, or network, for force conduction with a correlation length that can be decoupled from the actual particle size or system size \cite{gilabert2008,gaume2017,roy2016}.

As an alternate perspective, the stiffness of aggregated materials such as dilute colloidal gels \cite{shih1990,mewis2012} and more dense powder agglomerates \cite{kendall1987,kendall2001} is often instead described only in terms of the volume fraction $\phi$.  In these cases, power laws are also commonly seen, with $E\sim\phi^{\gamma}$ and where $3<\gamma<5$.  Again, the prevailing consensus is that this reflects the scaling of an effective backbone of stress-bearing elements \cite{mewis2012}.  

\subsection{Onset of failure}

As compression continues the stress-strain curves of Fig.~\ref{fig:stress-activity}(a) all show either a peak representing an ultimate compressive strength or a plateau indicative of a yield stress.  Generally, higher initial packing fractions can sustain higher stresses, and strains, before beginning to fail. This can be compared to experiments involving the uniaxial compression of porous alumina \cite{vives13a}, along with sandstones \cite{baud2014} and their corresponding DEM simulations \cite{schopfer2009,mcbeck2019}, all of which demonstrate that porosity exerts a strong control on the strength of cohesive granular materials.  There are several reasons for this.  First, the absolute size of pores will have a tendency to be larger in more porous materials (assuming that the grain size is roughly constant) and will thus act as larger flaws for the purposes of concentrating stress above any limiting fracture toughness \cite{sammis1986}.  Additionally, recent work on the transmission of force chains through cemented granulates suggests that there will be a greater localization of stress in more porous samples, significantly reducing their ultimate capacity to bear load \cite{mcbeck2019}.

In our case, the stress-strain curves for samples between $\phi=0.525$ and $0.580$ show clear peaks, corresponding to their brittle failure mode. On the other hand, for intermediate volume fractions the post-peak stress is maintained at a roughly constant level, which is consistent with a continuous yielding process, or plastic/ductile failure.  At even lower packing fractions, the stress-strain curves become more intermittent, showing irregular drops in stress and at the very lowest $\phi$ simulations were halted before any clear stress plateau emerged.   In the cases where it was reasonably well-defined (specifically, for $\phi\geq 0.475$) we measured how the peak stress, $\sigma_{\mathrm{p}}$, sustained during compression depends on the initial bond or contact density.  These data, shown in Fig.~\ref{fig:stress-activity}(c) can be fit by a power law of exponent $\alpha = 2.53\pm 0.12$.  Although expressed only over a small range of $Z\phi$, this is in good agreement with the power law scaling, of exponent $3.04$, reported by Gaume \textit{et al.} \cite{gaume2017} for the ultimate compressive strength of cohesive granulates made of sticky hard spheres.  

In what follows we will focus on characterizing the three types of failure that occur in our simulations, as highlighted in Fig.~\ref{fig:stress-activity}(d-f), and the transitions between these failure modes.  For example, these panels also show how the frequency of microcracks changes throughout each compression test. We measure this microcrack activity rate in terms of the number of broken bonds in each strain step, of size $3 \times 10^{-4}$.  In experiments the bond breakage rate can be estimated by monitoring acoustic emissions as a material sample deforms (\textit{e.g.} \cite{lockner1991,sammonds1992,olsson1999,baud2004,fortin2006,townend2008}). Figure~\ref{fig:stress-activity}(d) shows how, in simulations for $\phi=0.550$, a peak of microcrack activity occurs near the peak stress. The ductile response for $\phi=0.500$, shown in Fig.~\ref{fig:stress-activity}(e), exhibits fewer microcracks, and a less well-defined peak.  Finally, at the lowest packing densities abrupt spikes or bursts of activity accompany the intermittent stress drops.  This can be seen in Fig.~\ref{fig:stress-activity}(f) near $\epsilon\simeq 0.22$ and $0.28$, and is reminiscent of the intermittency of acoustic activity which accompanies compaction band formation in some more porous sandstones \cite{baud2004}, for example.   
In all cases, the spatial distribution of the internal damage changes near failure, as can be highlighted by a statistical analysis of the microcrack activity.  

\subsection{\label{sec:yielding_point}Entropy of microcrack activity near failure}

As the initial volume fraction of our simulated cohesive granular material changes, it responds to compression by very different deformation processes.   These reflect differences in the organization of damage in the samples during and beyond the elastic regime.  As an indicator of the spatial distribution of the microcracks or bond failure events, we calculated the normalized configuration entropy of the microcrack locations as compression proceeds. Any clustering of microcracks, or localization of damage, can be detected through a decrease in this entropy \cite{garcimartin1997}, just as the entropy of a gas goes down when its atoms cluster together, rather than randomly filling space. The widespread adoption of this metric also allows for comparisons across a broad range of experiments and simulations of mechanical deformation. In this context the configuration, information or Shannon entropy has been used to predict the failure point of brittle and inhomogeneous materials, for example in experiments involving plaster, wood and fiberglass \cite{garcimartin1997,guarino1998}.   A decrease in this entropy has also been reported to accompany catastrophic failure in field observations, for example in acoustic signals in a galena mine \cite{lu2005}.  Similarly, entropy measurement is used for  applications such as the analysis of generic random-fuse models \cite{reurings2005} or earthquake time series \cite{matcharashvili2002, telesca2004, padhy2004, de2011, bressan2017}. In all these cases the entropy, which describes whether damage is either spread out or localized and correlated throughout a sample, reduces towards the failure point.

We calculated the configuration entropy of the positions of microcrack activity following methods adapted from Refs. \cite{garcimartin1997,guarino1998}.  Specifically, by dividing the cylindrical system into $10\times 10\times 15$ cells, of size $2.2d\times2.2d\times2d$ (\textit{i.e.} slightly shorter than cubic cells), we measured the fraction of microcracks, $q_i$, occurring in each cell $i$ in any given strain window.  For these windows we divide the strain into 20 equal bins, each approximately 0.01 in width.  The normalized entropy is then calculated as
\begin{equation}
\label{eq:entropy}
    S=-\frac{1}{S_0}\sum_{i} q_i \ln q_i.
\end{equation} 
Empty cells, outside the cylinder, are excluded from this summation.  For ease of use, $S$ is normalized by the equipartition entropy, $S_0$, which is the entropy sum calculated by assuming that all the microcracks occurring in that strain window were instead completely randomly distributed in space.  This means that if bond breakage is indeed happening randomly, it would maximize the entropy with $S=1$. As irreversible damage becomes more localized, $S$ decreases, approaching $S=0$ for perfectly co-localized events. However, while any localization trend should be well-captured by variations of entropy, the absolute value can depend on the chosen cell sizes \cite{guarino1998}, which should like comfortably between the particle and system scales.   Although not tested here, the magnitude of $S$ will also likely vary with the system size, as the relative widths of shear bands in cohesive granulates are known to depend on the sample size \cite{pal2016}.

\begin{figure}
    \includegraphics[width=1.\linewidth]{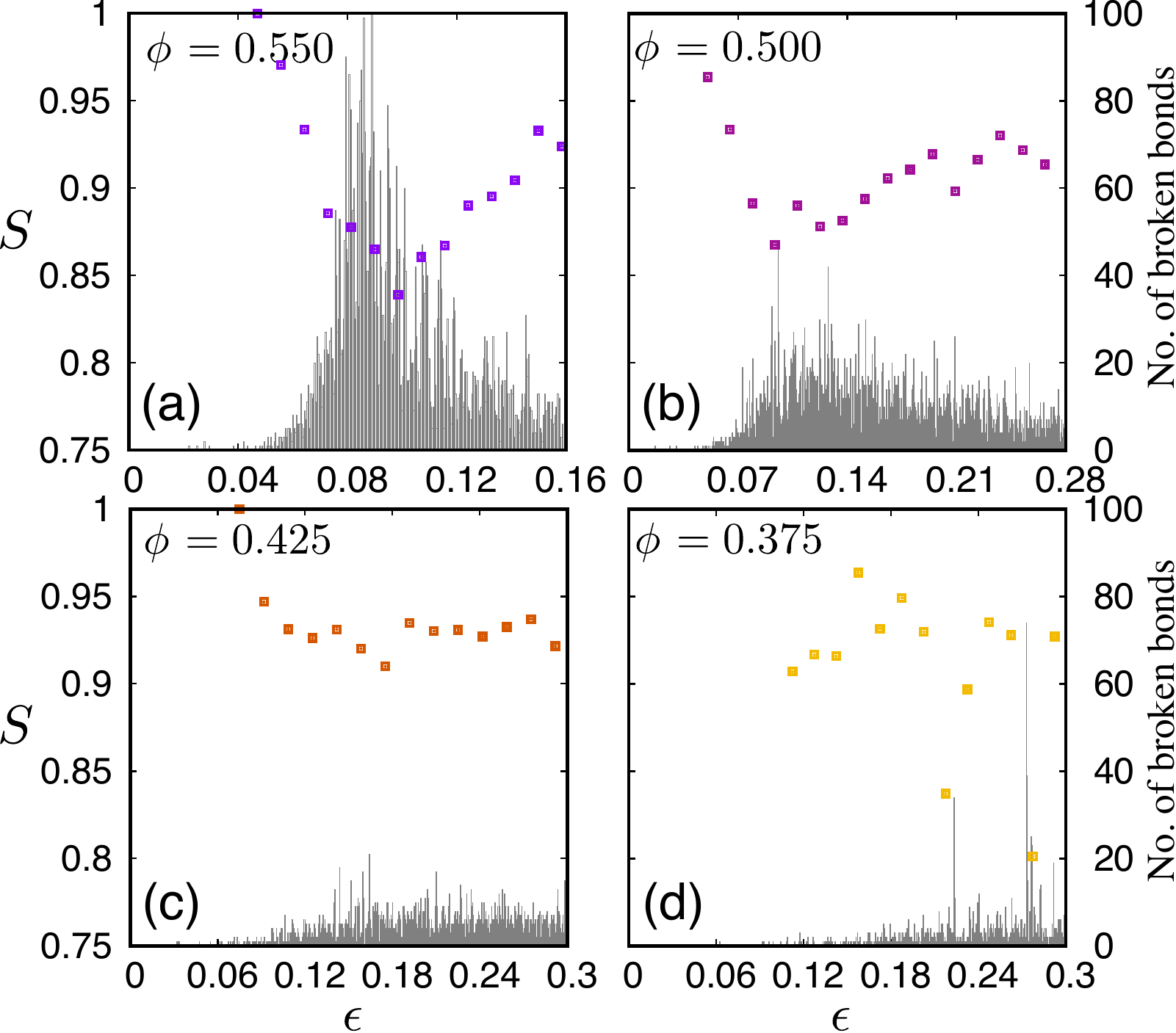}
    \caption{The normalized configuration entropy, $S$, characterizes the localization of damage during failure.  Shown here is the strain-dependence of $S$ and its associated microcrack activity for model realizations with initial volume fractions, $\phi$, of (a) $0.550$, (b) $0.500$, (c) $0.425$ and (d) $0.375$. Values of $S$ were calculated using Eq.~\ref{eq:entropy}, by binning bond failure events over regular strain intervals. At high $\phi$ the drop in $S$ is simultaneous with the increase in activity as a shear-band forms, and shows precursor activity before the stress peak (compare with Fig.~\ref{fig:stress-activity}(d)).  At intermediate $\phi$ the entropy drops to a lesser degree, and maintains a constant value as yielding proceeds.  At the lowest $\phi$ the intermittent spikes of activity, coinciding with stress drops, show low $S$ and hence highly localized damage.}
    \label{fig:entropy-activity}
\end{figure}

The variations of the normalized entropy, $S$, with strain are shown in Fig.~\ref{fig:entropy-activity} for various packing fractions, superimposed over the global microcrack activity (calculated as in Fig.~\ref{fig:stress-activity}). 
With an increase in strain, the damage is localized in space and the rate of bond breakages increases. The localization trend appears as a reduction of entropy in terms of the spatial distribution of bond breakages.

At high packing fractions, as in Fig.~\ref{fig:stress-activity}(a), the entropy begins to drop well before peak stress is reached, as bond breaking events become more common.  This precursor signal to shear-band formation is similar to that observed in typical experiments of fracture in heterogeneous brittle materials \cite{garcimartin1997,guarino1998}.  Additionally, we see that $S$ recovers to higher values after peak stress, as can be observed in earthquake time series after the mainshock and aftershocks \cite{bressan2017}, for example.  Here, however, we attribute this to a gradual widening of the shear band as irreversible damage accumulates; without a mechanism for healing the model system will never come to a steady-state slip condition.   In some simulations secondary bands also occur, as deformation proceeds.  Both mechanisms will serve to increase $S$, by adding to the volume over which strains and damage are focused.  
As the packing fraction decreases, the drop in entropy becomes less pronounced.  As shown in Fig.~\ref{fig:entropy-activity}(b), for $\phi = 0.500$ there is still a noticeable dip in $S$ leading up to and accompanying the peak stress and peak microcrack activity.  However, for $\phi=0.425$, as in Fig.~\ref{fig:entropy-activity}(c), the entropy maintains a constant value throughout the yielding process.  Inspection of where the damage is happening, for example as can also be seen in Fig.~\ref{fig:snapshot}(b,c), suggests that the weaker depression of $S$ that still occurs in such cases is due to damage accumulating near the upper and lower boundaries, as a result of the clamped boundary conditions used.  

Perhaps most intriguingly, for the lowest packing fractions we see intermittent fluctuations in $S$ that accompany the formation of anti-cracks.   Here, where there is a drop in stress and a spike in bond breakage activity, the corresponding entropy drop records a shift from a pattern of random damage accumulation to a very localized phenomenon.  For the case of $\phi=0.375$ this can be seen in Fig.~\ref{fig:entropy-activity}(d), near $\epsilon = 0.22$ and $0.28$, and this data is typical of our low-$\phi$ samples.   We are not aware of any experiments that directly quantify $S$ during anti-cracking.  However, comparison of our results to reports of the localization of compaction in acoustic emission studies of sandstone \cite{olsson2000}, or which visualize compaction band growth in sandstones \cite{baud2004} snow \cite{barraclough2017} and foam \cite{reis2009} suggest that the intermittent fluctuation of the positional entropy of microcrack activity may be a general feature of compaction band or anti-crack formation. 

\subsection{\label{sec:failure_process} Transitions between failure process}

In our simulations we have identified three distinct failure modes, which include (i) brittle failure with shear banding, (ii) ductile failure with plastic deformation and (iii) compaction with the formation of anti-crack, as was detailed in Fig.~\ref{fig:snapshot}.   We will here show how these modes can be distinguished in terms of the evolution of their local packing fractions, the spatial profiles of their beads' coordination numbers and displacements, and by variations in the positional and angular correlations of microcracks.

\begin{figure}
    \includegraphics[width=1.\linewidth]{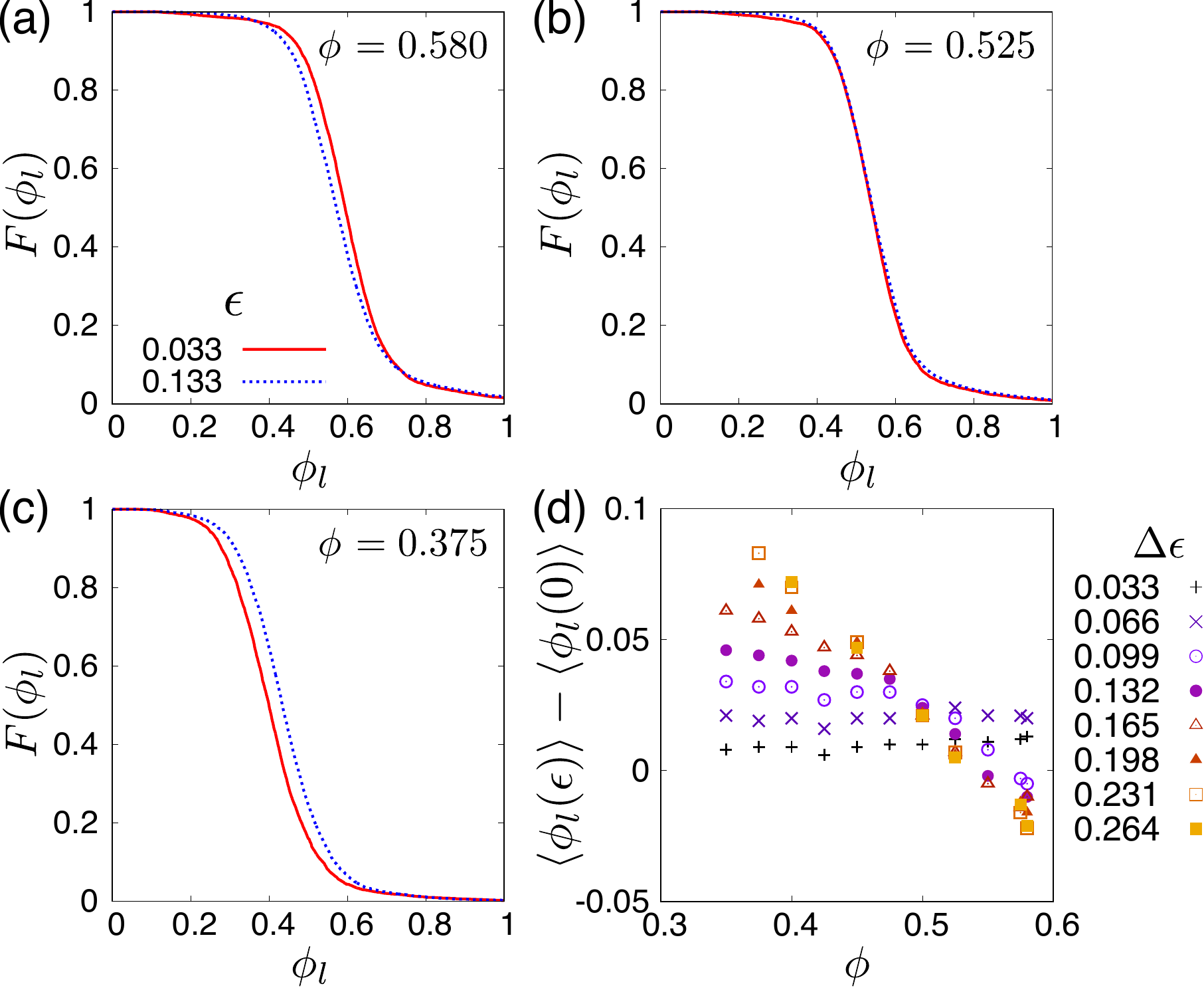}
    \caption{Simulations demonstrate dilatancy at large enough $\phi$, but not for more porous materials.  We show here the complementary cumulative distribution function of the \textit{local} packing fraction for (a) $\phi=0.580$, (b) $0.525$ and (c) $0.375$ and strains of $\epsilon=0.033$ (red solid lines) and $0.133$ (blue dotted lines). The shift of the distribution from left to right with an increase of strain implies the compaction of a system. The shift for the opposite direction implies dilation. (d) At various strains we can identify the median $\phi_l$ as the the half-way point of such curves, \textit{i.e.} where $F=0.5$.  Plotting how this median, $\langle \phi_l (\epsilon) \rangle$, changes from its initial value,  $\langle \phi_l (0)\rangle$, highlights the dilatancy.  Here, a positive dilatancy factor is implied for $\phi = 0.525$ and above, which is also where shear bands are seen.}
    \label{fig:cumulative_local_phi}
\end{figure}

Beginning with considerations of local packing fractions and dilatancy, we calculated the cumulative distribution function of the local packing fractions after different strains, using the VORO$++$ code library \cite{rycroft2009}. The local packing fraction of particle $i$ is defined by $\phi_l={V_i}/{\tilde{V}_i}$, where $V_i=(\pi/6) D_i^3$ and $\tilde{V}_i$ is the volume of its Voronoi cell (\textit{e.g.} as in Refs. \cite{voronoi1908,aste2007}). The cumulative distribution function is then $F(\phi_l)=1-\int_0^{\phi_l} P(\phi) d\phi$, for the probability distribution function $P(\phi)$. The results are shown in Fig.~\ref{fig:cumulative_local_phi}. 

For high packing fractions, such as the case of $\phi=0.580$ shown in Fig.~\ref{fig:cumulative_local_phi}(a), the distribution function $F(\phi_l)$ shifts to higher volume fractions as compression proceeds. This is a smooth increase, occurring at all volume fractions.  A similar effect is seen in the motivating experiments, in which $\phi_l$ can be calculated from X-ray tomograms, and which show a reversible Reynolds dilatancy \cite{hemmerle2016,hemmerle2020}.  In the simulations the extent of dilatancy reduces rapidly as the material becomes more porous and is barely noticeable by $\phi = 0.525$, as demonstrated in Fig.~\ref{fig:cumulative_local_phi}(b).  In contrast, for low $\phi$ the material compresses more intuitively, with all particles reducing their local packing fraction, on average.  This case is shown in Fig.~\ref{fig:cumulative_local_phi}(c), for $\phi=0.375$.  We can conclude that there is the cross-over between dilation and compaction with decreasing $\phi$, as seen in terms of the cumulative distribution functions of the local packing fractions. 

To better illustrate this cross-over point, we also measured how the median of the local volume fraction changes with increasing strain.  This is found by noting that the median value, $\langle \phi_l \rangle$, occurs where $F=0.5$.  Figure~\ref{fig:cumulative_local_phi}(d) shows how $\langle \phi_l \rangle$ changes away from its initial value, as samples with different initial packing fractions are compressed.   We find that the change from a dilatant response, where $d \langle \phi_l \rangle / d\epsilon < 0$, to a compressive response, where $d\langle \phi_l \rangle / d\epsilon > 0$ happens between $\phi = 0.500$ and 0.525.  Interestingly, this cross-over coincides with the change from brittle failure, \textit{via} shear band formation, to a plastic failure.  This condition is also consistent with a change in the sign of the dilatancy factor, $\beta$, which in geomechanics derives from the ratio of the plastic volumetric to axial strains under deformation \cite{wong2012,baud2005}.  In good agreement with experiments, when sandstones are compressed clear shear bands are seen for $\beta>0$, whereas low-angle bands, compaction bands or homogeneous cataclastic flows are seen for $\beta<0$ \cite{wong2001,baud2005}.   

\begin{figure}
    \includegraphics[width=1.\linewidth]{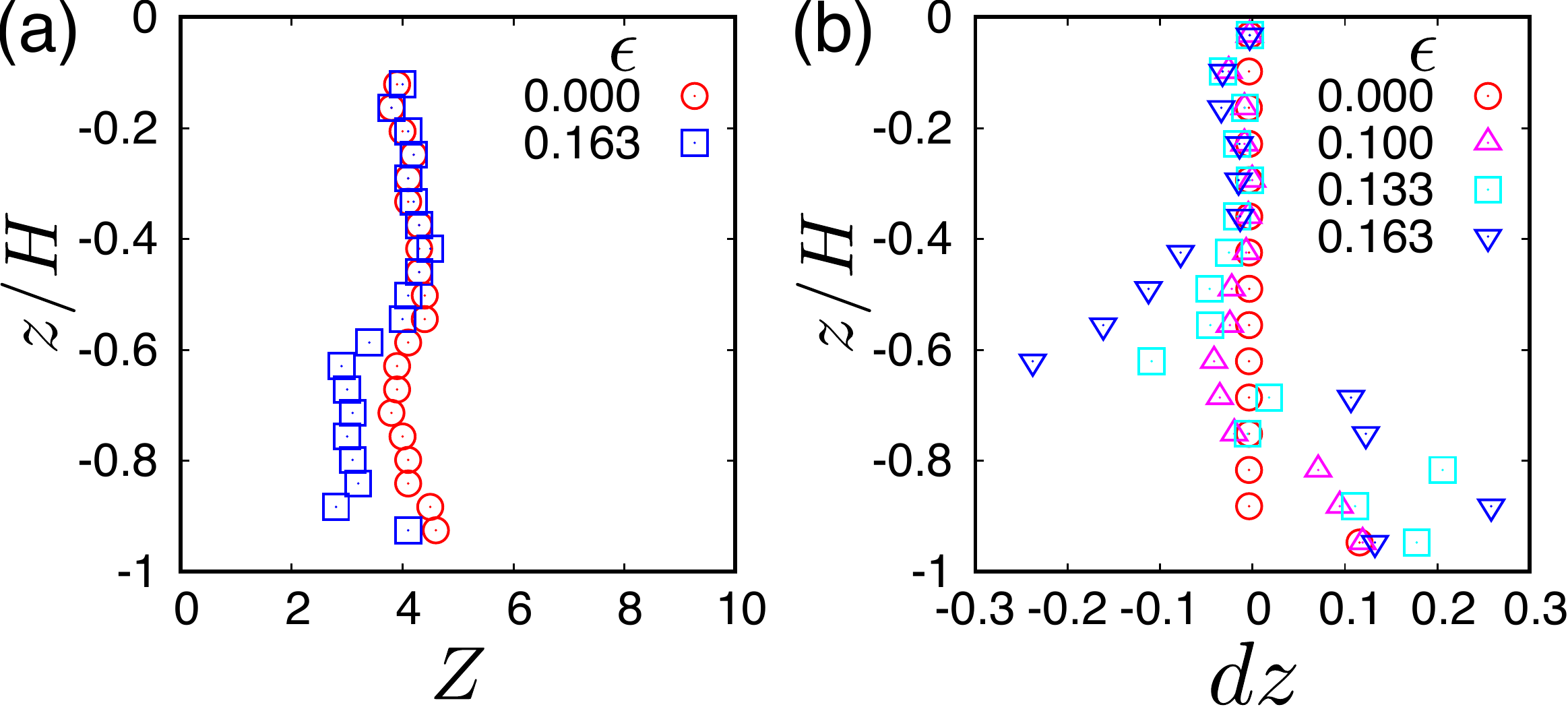}
    \caption{Detection of an anti-crack at $\phi = 0.400$. We show here how the horizontally-averaged (a) coordination number $Z$ and (b) vertical displacement $dz$ of particles in a sample changes under compression.  In both panels the positional information is plotted using a Lagrangian or undeformed reference frame (as in \textit{e.g.} \cite{reis2009} for anti-cracks in foam), to allow features to be more easily compared.  At a depth of between 0.6 and 0.7 times the sample height an anti-crack forms.  This can be seen by a local reduction of the coordination number and by the opposite signs of the $z$-displacement profile above and below the anti-crack.  
    }
    \label{fig:anti-crack}
\end{figure}

Next, further detailing the anti-crack formation seen at lower packing fractions, we describe how the coordination number and the vertical displacement of particles in a typical porous sample evolve during compression.   Anti-cracks are a compressional instability with a displacement profile similar to an opening (mode-I) crack, but with an opposite sign \cite{fletcher1981,sternlof2005,heierli2008,reis2009}.  They can be identified by localized reductions in porosity \cite{sternlof2005}, microcrack activity \cite{townend2008}, damage accumulation \cite{wong2001}, strains and displacements \cite{reis2009}.  Here, we characterize the damage localization of these features in our simulations by plotting the horizontally-averaged coordination number of a simulation of $\phi=0.400$; the data are taken from the simulation shown in Fig.~\ref{fig:snapshot}(c), where an anti-crack can be seen to develop from one side of the sample to the other between strains of 0.11 and 0.16. As shown in Fig.~\ref{fig:anti-crack}(a), the anti-crack reduces the average number of contacts in a  area about two-thirds of the way down the cylinder.  At the same position, Fig.~\ref{fig:anti-crack}(b) shows that there is a discontinuity in the vertical displacement of the particles.  In that panel we have calculated the horizontally-averaged displacement of the particles, after subtracting out the expected uniform compression.  In both cases we have plotted positions in an undeformed coordinate system, normalized by the original sample height, $H$.

\begin{figure}
    \includegraphics[width=1.\linewidth]{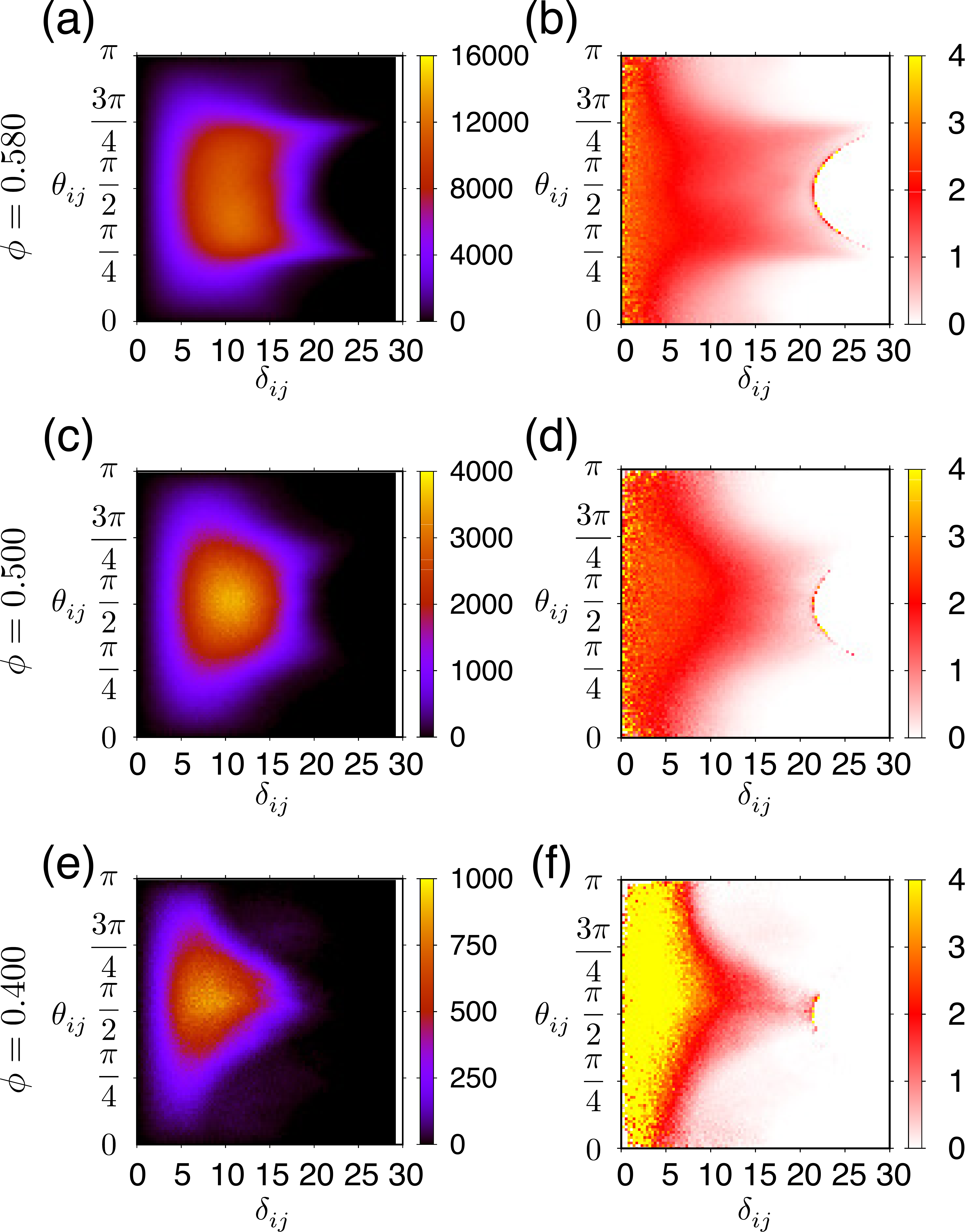}
    \caption{Correlation analysis of damage in various modes of failure.  Shown on the left are density maps giving the relative likelihood of finding pairs of bond breakages at distances $\delta_{ij}$ and angles $\theta_{ij}$ from each other.  
    Panels on the right show the same information, normalized by a similar map generated from pairs of points randomly distributed within the volume of the sample.  In this representation, bond breakages happening entirely at random would give a correlation value of 1.  Note that the cylindrical shape of the samples constrains the choices of allowed pairs of bonds and particles, and is responsible for the curve of noise around $\delta_{ij} = 20$. Here, panels (a) and (b) show the brittle failure at $\phi=0.580$, (c) and (d) are at $\phi=0.500$ and highlight ductile deformation, while (e) and (f) present data from an anti-crack at $\phi=0.400$.}
    \label{fig:density_map}
\end{figure}

Finally, we show how the different patterns of damage in the various failure modes seen can also be well-described by a correlation analysis, based on the spatial distribution of bond breakage events.  To calculate this correlation function we considered the locations of all such events throughout a numerical experiment.  We then found the displacement vectors between all possible pairs of broken bonds.  If averaged over all orientations, this would allow calculation of the pair correlation function $g(r)$.  However, to capture the difference between shear and compaction bands, we measured the length $\delta_{ij}$ and polar angle $\theta_{ij}$ (\textit{i.e.} angle measured from the axis of compression) of these pair-wise displacement vectors.  In Fig.~\ref{fig:density_map} we present results from compression tests calculated in this way.  The right hand-side of this figure shows the same data, normalized by a correlation function calculated in a like manner but for random positions within the cylindrical sample.  Figures~\ref{fig:density_map}(a,b) show the case of a shear band.  Here, the inclination of the band at $45^\circ$ to the direction of compression is seen in the strong correlations at $\theta_{ij} = \pi/4$ and $3\pi/4$; this implies the localization of microcracks along diagonal planes.   The case of diffuse plastic damage, given in Fig.~\ref{fig:density_map}(c,d), shows weaker correlations in any orientations.   The last case, of an anti-crack, is given in Fig.~\ref{fig:density_map}(e,f), and shows the localized events in the horizontal direction of $\theta_{ij}=\pi/2$.  Here, the width of the anti-crack feature, of about 5 particle diameters, is apparent by the increased correlations of bond breakages below this scale. 

\section{\label{sec:discussion}Summary and conclusions}

In this paper, we have modeled a cemented or cohesive granular system by means of discrete element model (DEM) simulations.  The simple yet fully constrained model can correctly predict considerable detail of the various modes of failure and elastic responses, and is therefore a  strong candidate for venturing into situations that are hard to yet probe experimentally.

More specifically, the parameters in our model are carefully chosen so as to be consistent with experiments \cite{hemmerle2016, schmeink2017, hemmerle2020} involving cohesive materials constructed by glass beads connected by relatively compliant polymer bridges.   Its interactions include an energy-based failure criterion for the bonds between cemented or aggregated particles and bonds which themselves take experimentally measured spring constants.   Despite the simplicity of the inputs, the model accurately reproduces the behaviors of its experimental analogue under compression.  It has a similar stress-strain curve, elastic modulus, fails at appropriately large strains by shear failure, and shows dilatant behavior prior to failure, for example. 

By varying the packing fraction of the particles and the geometry of their connections, while keeping all other interactions constant, we showed how our model was also able to reproduce other phenomena associated with the mechanical deformation of cohesive granular media.  This includes details such as the particularly strong power-law scaling of the elastic modulus and compressive strength of cohesive granular materials with the density of bonds (as with \cite{gaume2017}), effectively described by $Z\phi$.  These results, in particular, lend support to the developing idea of how force chains control the failure processes in cohesive granular systems \cite{gilabert2008,baud2014,roy2016,mcbeck2019}, by controlling the distribution and density of stress concentrations.  

The model also provides an excellent framework for exploring transitions between failure modes, as it can reproduce shear banding, plastic creep and anti-cracks or compaction banding, in one unified system.  These three types of failure are typical of porous materials including snow \cite{kinosita1967,heierli2008,barraclough2017}, foam \cite{reis2009}, colloidal gels \cite{shih1990,roy2016,mewis2012}, powder aggregates \cite{kendall1987,kendall2001,gilabert2008,vives13a}, sandstone \cite{baud2004,baud2005,fortin2006,baud2012,eichhubl2010,schultz2010,fossen2011,townend2008} and so on.  As an example, we demonstrated how the shift in the sign of the dilatancy factor corresponds to a shift from brittle to plastic failure, in a manner very similar to what has been reported in compressional tests on sandstone \cite{baud2005}. Additionally, we showed how anti-crack formation is associated with intermittent stress drops, strain and damage localization, and is accompanied by temporary reductions in the positional entropy of microcrack activity.

There remains wide scope for further application of this model.  Its parameters can be changed to reflect different materials, from snow (low $\phi$, weak bonds) to sandstone (high $\phi$, stiff bonds with strength depending on cement or matrix content) and artificial composites like the materials that directly inspired it.  The boundaries between failure types could be explored further, and system size effects investigated, as was done for shear band width in Ref. \cite{pal2016}.  For example, we intend to extend the comparisons to our experiments to include a study of the effects of bond stiffness \cite{hemmerle2020}.  A similar effort could explore the effects of varying the volume fraction of the bond material, to allow for a broader characterization of rocks, such as in \cite{wang2019,delenne2009}.  

\begin{acknowledgments}
This work was supported by the Leading Graduate Course for Frontiers of Mathematical Sciences and Physics, The University of Tokyo, MEXT. The authors would like to thank Arnaud Hemmerle (Synchrotron SOLEIL) for offering the experimental data and helpful discussions. Y. Y. and T. H. also acknowledge Grants-in-Aid for Scientific Research (KAKENHI) No. JP16H06478.
\end{acknowledgments}

\appendix
\nocite{*}

%

\end{document}